\documentclass[conference]{IEEEtran}

\usepackage[T1]{fontenc}
\usepackage[utf8]{inputenc}
\usepackage{cite}
\usepackage{url}
\usepackage{booktabs}
\usepackage{array}
\usepackage{tabularx}
\usepackage{multirow}
\usepackage{enumitem}
\usepackage{xcolor}
\usepackage{hyperref}
\usepackage{graphicx}

\usepackage{tabularx}
\usepackage{multirow}
\usepackage{colortbl}
\usepackage{longtable}

\hypersetup{
    colorlinks=true,
    linkcolor=black,
    citecolor=black,
    urlcolor=blue
}

\begin{document}

\title{Operationalizing Responsible AI in MLOps and ML Production: A Systematic Literature Review}

\author{

\IEEEauthorblockN{1\textsuperscript{st} Hinduja Nirujan}
\IEEEauthorblockA{\textit{Electrical and Software Engineering} \\
\textit{University of Calgary}\\
Calgary, Alberta, Canada \\
hinduja.nirujan@ucalgary.ca}
\and
\IEEEauthorblockN{2\textsuperscript{nd} Siddhartha Paudel}
\IEEEauthorblockA{\textit{Electrical and Software Engineering} \\
\textit{University of Calgary}\\
Calgary, Alberta, Canada \\
siddhartha.paudel1@ucalgary.ca}
\and
\IEEEauthorblockN{3\textsuperscript{rd} Gouri Ginde}
\IEEEauthorblockA{\textit{Electrical and Software Engineering} \\
\textit{University of Calgary}\\
Calgary, Alberta, Canada \\
gouri.ginde@ucalgary.ca}
}

\maketitle

\begin{abstract}
Artificial intelligence (AI)-enabled systems are increasingly deployed in operational environments where machine learning (ML) models interact with changing data, users, infrastructure, and organizational processes. While Responsible AI principles are widely discussed, less is known about how these principles are translated into concrete mechanisms within Machine Learning Operations (MLOps) and ML production workflows. This paper reports a systematic literature review of studies that operationalize Responsible AI in ML production. The review synthesizes evidence across monitoring and runtime assurance, audit and compliance, ML lifecycle integration, and traceability/provenance/documentation. Across the final study set, the literature shows a shift from abstract ethical principles toward engineering mechanisms such as drift monitoring, runtime checks, audit logs, provenance records, evidence bundles, governance gates, policy-as-code, model passports, and compliance-oriented documentation. However, the evidence base remains uneven: many studies propose architectures or prototypes, while fewer evaluate the mechanisms in realistic production settings, longitudinal operations, or organizational governance workflows. The review identifies a central gap around integrated, evidence-producing Responsible MLOps: current mechanisms often detect, document, or govern isolated parts of the ML lifecycle, but rarely connect runtime signals, ML lifecycle provenance, human accountability, and compliance evidence into closed operational loops. These findings motivate future software engineering research on traceable, evaluable, and operationally integrated Responsible AI support for ML production.
\end{abstract}


\section{Introduction}

Concerns around Artificial Intelligence (AI) ethics have expanded as the adoption of AI in software systems has increased significantly in recent years. Sectors such as healthcare, finance, business, education, security, and justice are already incorporating AI-based systems through Machine Learning (ML) and Deep Learning algorithms \cite{ Putting_ai_ethics_work, SR_Ethics_tools_2024,policy_to_pipeline_2026}. In response, multiple frameworks, principles, guidelines, and consultation processes have been proposed by institutions like governments, international bodies, civil society organizations, industry, and research to support the development of trustworthy and responsible AI systems \cite{RAI_Patterns_SE}. However, AI-enabled software systems differ from traditional software systems because they involve interdependent activities such as software engineering (SE), data engineering, data science, model development, deployment, and maintenance. The engineering and operational management of ML-related services in software systems are commonly addressed through Machine Learning Operations (MLOps), which extends DevOps concepts to support the lifecycle of ML-enabled systems \cite{MLOps_architecture}.

AI-enabled systems deployed in production environments can influence decisions, services, and organizational processes \cite{MLOps_adoption_RAI, MLR_MLOps_practices, data_centric_AI_governance_2026, AI_driven_retail_decision_2026}. As these systems move from experimental development to operational use, reliability can no longer be treated only as a design-time concern \cite{RAI_pattern_catalogue_2024,semantic_AI_System_accountability_2021}. 

Earlier, AI governance has mainly relied on high-level ethical principles, frameworks, and documentation checklists \cite{data_cards_2022, DocML_2023}. These approaches helped define important Responsible AI values, such as fairness, explainability, safety, transparency, and accountability, but provided limited guidance on how engineers should implement, verify, and audit these requirements in operational systems \cite{policy_as_code_mining_2026,role_of_XAI_healthcare_2022}. This gap has become more important as AI regulation has developed rapidly. Regulations such as the European Union Artificial Intelligence Act (EU AI Act), the General Data Protection Regulation (GDPR), and the European Digital Identity Framework (eIDAS 2.0) increasingly require organizations to provide clear and auditable evidence of technical controls \cite{serverless_covariate_drift_2024, xai_compliance_2026}. For high-risk AI systems, this includes demonstrating effective risk management, continuous logging and record-keeping, robustness to unexpected inputs or conditions, and meaningful human oversight \cite{Ergo_smirk_2023, compliance_as_code_2026}.

However, verifying and assuring ML-based software systems is more difficult than for traditional software. Traditional safety standards, such as EN 50126-1 and IEC 62443, are largely based on deterministic system behavior, static code reviews, and predefined test coverage\cite{policy_as_code_mining_2026}. ML systems differ because their behavior depends heavily on data. They may be non-deterministic, and can degrade after deployment due to concept drift or covariate drift\cite{serverless_covariate_drift_2024, augur_2023, hybridMLOps_2025}. When the data distribution changes, model performance may decline without producing obvious system failures, while standard operational metrics such as CPU and memory usage remain normal. As a result, trust in an ML system cannot be established only before deployment\cite{hybridMLOps_2025}. It must be continuously monitored and assessed throughout the system lifecycle using automated and machine-readable evidence collected from the operational pipeline.

\subsection{The importance of monitoring and auditability}
To operationalize RAI in production pipelines, this Systematic Literature Review (SLR) focuses on two main pillars: monitoring and auditability.
\begin{enumerate}[leftmargin=*]
\item Monitoring:\\ 
Monitoring focuses on detecting runtime anomalies, data distribution shifts, adversarial inputs, and demographic bias before they negatively affect system behavior\cite{monitoring_practices_2026}.
\item Auditability:\\
Auditability supports accountability by collecting and preserving machine-readable evidence. This may include dataset provenance, training configurations, software bills of materials (SBOMs), validation results, and records of runtime decisions\cite{semantic_AI_System_accountability_2021, policy_to_pipeline_2026, compliance_as_code_2026}.
\end{enumerate}

In modern MLOps workflows, these two pillars work together to create an automated safety and compliance loop. As a model processes incoming data, separate monitoring components can evaluate data quality and distribution without slowing down the prediction process. These monitoring signals provide important evidence for continuous auditing. Instead of relying on manual and time-consuming documentation, modern architectures increasingly use Compliance-as-Code (CaC). In this approach, legal, safety, and ethical requirements are translated into machine-readable rules that can be evaluated automatically using policy engines such as Open Policy Agent (OPA)\cite{policy_to_pipeline_2026, compliance_as_code_2026}. These rules check system configurations and monitoring results at important stages of the deployment pipeline. If a model does not meet required safety thresholds or does not have verifiable provenance, the pipeline can automatically block the update from reaching production. The results of these checks can then be exported in standardized, machine-readable formats, such as the Open Security Controls Assessment Language (OSCAL), creating a clear and traceable link between technical evidence and the original regulatory requirements\cite{policy_to_pipeline_2026}.

Although academic and industry research has proposed advanced approaches, their use in real-world ML systems remains limited. To examine this gap, this systematic literature review combines a literature analysis of 44 primary studies that investigate how Responsible AI principles are operationalized within MLOps and ML production workflows. The review examines the technical mechanisms, architectures, governance controls, and operational challenges reported in the literature, with particular attention to the gap between proposed Responsible AI solutions and their practical adoption.

To examine this gap between regulatory goals and real-world engineering practice, this SLR is organized around three Research Questions (RQs). These questions follow the progress from monitoring and evidence collection to governance enforcement and, finally, practical adoption challenges.
\begin{itemize}
    \item RQ1: What technical architectures, statistical methods, and mechanisms are used in MLOps pipelines to monitor model behavior at runtime and maintain reliable dataset and model lineage?
    \item RQ2: How are legal, ethical, and safety requirements translated into executable rules and pipeline-level controls that govern model deployment throughout the ML lifecycle?
    \item RQ3: What operational, behavioral, and structural challenges limit the adoption of trustworthy MLOps in practice?
\end{itemize}

The remainder of this paper is organized as follows. Section \ref{methodology} describes the study selection process, inclusion and exclusion criteria, and data extraction and synthesis methodology applied to the 44 primary studies. Section \ref{results} explains the findings from the data extraction. Section \ref{RQ1} (RQ1: Observability and Traceability) examines approaches for drift monitoring, root-cause analysis, and dataset and model lineage tracking. Section \ref{RQ2} (RQ2: Codified Governance and Continuous Assurance) reviews how regulatory and safety requirements are translated into executable controls. Section \ref{RQ3} (RQ3: Challenges of adoption) synthesizes findings from the reviewed literature to examine operational and socio-technical barriers, including limited Responsible AI logging, documentation challenges, governance integration difficulties, and verification limitations in LLM-based systems. Section \ref{future} discusses future research directions, and Section \ref{threats} addresses threats to validity. Finally, Section \ref{conclusion} concludes the paper.

\section{Methodology}
\label{methodology}

This SLR follows an established software engineering SLR guidance, including explicit research questions, search strategy, inclusion and exclusion criteria, staged screening, data extraction, and synthesis from \cite{kitchenham2007guidelines,page2021prisma}.

\subsection{Search Strategy}
The search strategy was developed to find literature where the concepts of Responsible AI, MLOps, and operational governance intersect. Therefore, we constructed the search string combining three concept groups: Responsible AI terminology, MLOps/lifecycle terminology, and operationalization mechanisms such as monitoring, auditability, traceability, provenance, logging, evidence, and compliance checking. 
The refined search string used in the review was:

\begin{quote}
\small
(\texttt{"responsible AI"} OR \texttt{"trustworthy AI"} OR \texttt{"ethical AI"} OR \texttt{"AI ethics"} OR \texttt{"AI governance"} OR \texttt{"AI accountability"} OR \texttt{"AI assurance"} OR \texttt{"responsible machine learning"} OR \texttt{"trustworthy machine learning"}) AND
(\texttt{"MLOps"} OR \texttt{"machine learning operations"} OR \texttt{"ML production"} OR \texttt{"production machine learning"} OR \texttt{"machine learning pipeline"} OR \texttt{"ML pipeline"} OR \texttt{"machine learning lifecycle"} OR \texttt{"model lifecycle"} OR \texttt{"model deployment"} OR \texttt{"ML deployment"} OR \texttt{"model serving"} OR \texttt{"model maintenance"} OR \texttt{"software engineering"} OR \texttt{"AI engineering"} OR \texttt{"software engineering for AI"} OR \texttt{"software lifecycle"} OR \texttt{"system development lifecycle"} OR \texttt{"SDLC"}) AND
(\texttt{"monitoring"} OR \texttt{"model monitoring"} OR \texttt{"runtime monitoring"} OR \texttt{"continuous monitoring"} OR \texttt{"post-deployment monitoring"} OR \texttt{"production monitoring"} OR \texttt{"drift detection"} OR \texttt{"concept drift"} OR \texttt{"data drift"} OR \texttt{"model drift"} OR \texttt{"auditability"} OR \texttt{"audit"} OR \texttt{"audit trail"} OR \texttt{"traceability"} OR \texttt{"provenance"} OR \texttt{"lineage"} OR \texttt{"logging"} OR \texttt{"telemetry"} OR \texttt{"observability"} OR \texttt{"compliance checking"}).
\end{quote}

 The search results were exported to Covidence for deduplication and screening \cite{covidence}. The initial database search was complemented by backward and forward snowballing from a strict subset of highly relevant seed studies, following snowballing recommendations for systematic reviews~\cite{wohlin2014snowballing}. This was used to identify recent or closely related studies that may not have been captured by the database search due to terminology variation.

\subsection{Selection Process}
\label{selection}
\begin{table}[t]
\centering
\caption{Inclusion and exclusion criteria used for study selection.}
\label{tab:criteria}
\begin{tabularx}{\columnwidth}{p{0.13\columnwidth} X}
\toprule
\textbf{Code} & \textbf{Criterion} \\
\midrule
\multicolumn{2}{l}{\textbf{Inclusion criteria}} \\
\midrule
IC1 & Addresses at least one Responsible AI concern, such as fairness, accountability, transparency, explainability, privacy, safety, robustness, compliance, auditability, or human oversight\cite{SR_Ethics_tools_2024}.\\

IC2 & Is situated in an ML lifecycle or production context, such as MLOps, deployment, serving, monitoring, maintenance, retraining, or software engineering for AI. \\

IC3 & Provides an operational contribution, such as a mechanism, artifact, architecture, framework, method, process, or tool for implementing or governing Responsible AI. \\

IC4 & Provides enough technical, methodological, or empirical detail to support replication of system design, evaluation, dataset, metrics, or evidence artifacts. \\

IC5 & Is available as full text in English and is peer-reviewed, or is a high-relevance preprint/conference version identified through snowballing or recent-venue search. \\

\midrule
\multicolumn{2}{l}{\textbf{Exclusion criteria}} \\
\midrule
EC1 & Discusses Responsible AI, ethics, trustworthiness, or governance only conceptually, without operationalization. \\

EC2 & Focuses only on model design, prediction accuracy, or benchmark performance without Responsible AI or ML lifecycle relevance. \\

EC3 & Focuses on general software engineering, DevOps, cybersecurity, or data management without clear AI/ML relevance. \\

EC4 & Is a broad survey, opinion, editorial, position, or vision paper without primary evidence of an operational framework. \\

EC5 & Lacks sufficient detail for extraction, has unavailable/incomplete full text, is abstract-only, or is not in English. \\

\bottomrule
\end{tabularx}
\end{table}
Table~\ref{tab:criteria} presents the inclusion and exclusion criteria used during study selection. The criteria were intentionally strict because the search topic spans several broad and overlapping areas, including Responsible AI, AI governance, MLOps, software engineering for AI, monitoring, auditability, and compliance. To be included, a study had to satisfy three core conditions: it had to address a Responsible AI concern, it had to be connected to ML production or ML lifecycle practice, and it had to provide an operational mechanism, artifact, process, framework, method, or tool. This requirement prevented the review from including papers that only discussed Responsible AI as a high-level principle without showing how it is operationalized.

During full-text screening, particular attention was given to whether the study provided extractable technical or methodological evidence. Studies were prioritized based on implementation details, ML lifecycle integration, evaluation design, datasets, metrics, practitioner evidence, deployment context, or governance artifacts. Papers that focused only on model accuracy, domain-specific AI application deployment, generic ethical principles, or regulatory discussion were excluded unless they clearly explained how Responsible AI was implemented or evaluated within an ML lifecycle setting.

The exclusion criteria were also applied conservatively during snowballing and recent-venue screening. Snowballing candidates were not included solely because they cited or were cited by a relevant seed paper. They were required to satisfy the same conceptual boundary as the main database search. This ensured that the final corpus remained focused on operational Responsible AI in MLOps and ML production rather than expanding into the broader literature on AI ethics, AI applications, or general software engineering. 

The study selection process consisted of title and abstract screening, full-text screening, and snowballing search. Screening was performed in Covidence by two authors.  The title and abstract screening showed a Cohen's kappa agreement score of 0.43. A total of 455 records proceeded to full-text screening. At this stage, the initial reviewer agreement was low, with a Cohen's kappa score of 0.28, indicating substantial ambiguity in applying the eligibility criteria to borderline studies. Conflicts were resolved through discussion at each stage, with final decisions guided by the inclusion and exclusion criteria described in Table~\ref{tab:criteria}.

\subsection{Selection Outcomes}

 \begin{figure}[t]
    \centering
    \includegraphics[width=\linewidth]{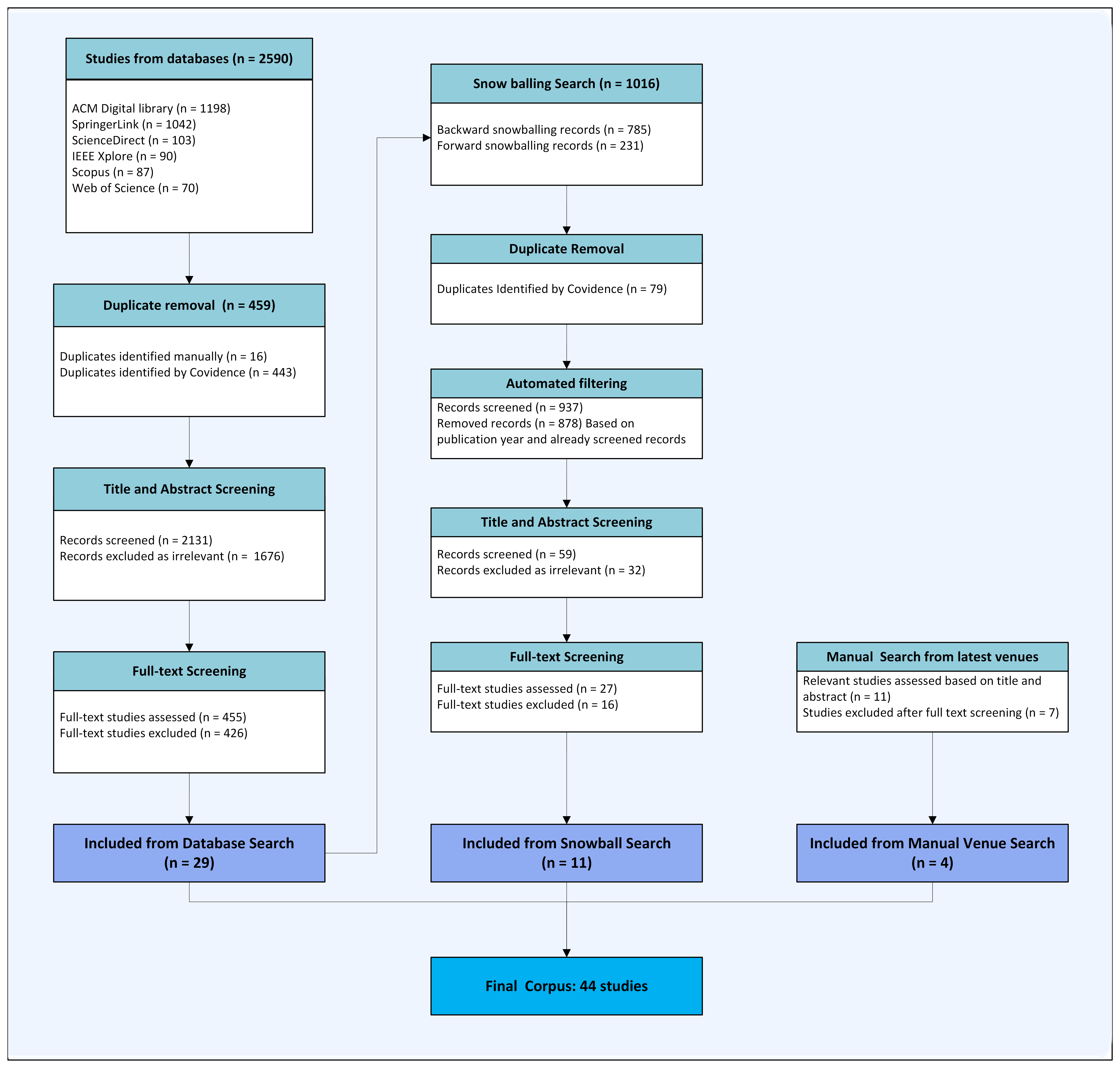}
    \caption{Study identification, screening, snowballing, and final synthesis selection process.}
    \label{fig:study-selection}
\end{figure}
Figure~\ref{fig:study-selection} summarizes the study identification and selection process. The review process consisted of database searching, duplicate removal, title and abstract screening, full-text eligibility assessment, supplementary snowballing, manual recent-venue checking, and final studies selection.

The database search identified 2,590 records across six sources: ACM Digital Library, SpringerLink, ScienceDirect, IEEE Xplore, Scopus, and Web of Science. After removing 459 duplicates, 2,131 records were screened by title and abstract. 1,676 records of these were excluded as irrelevant. The remaining 455 studies were assessed at full-text level, and 426 were excluded using the predefined exclusion criteria. Because the review topic is broad and many studies discuss Responsible AI only conceptually, a strict selection criteria explained in Section \ref{selection} was applied during the screening based on technical relevance, connection to MLOps or ML production, and evidence maturity. This resulted in 29 studies selected from the main database search for final corpus.

To reduce the risk of missing relevant studies, a round of backward and forward snowballing was conducted from the selected seed studies using OpenAlex \cite{openalex}. The snowballing search retrieved 1,016 records, consisting of 785 backward references and 231 forward citations. After deduplication, 937 candidates remained. Since snowballing was used as a supplementary search rather than a second exhaustive database search, a controlled filter was applied. This filter used the publication year range, and removal of records already screened during primary database search. The filtering step retained 59 records which were screened by title and abstract. In this step, 32 were excluded as irrelevant, 27 were assessed at full-text level. Finally, 16 were excluded during this stage and the remaining 11 studies were added to final corpus.

Finally, a targeted manual venue search was conducted to identify recent studies that may not yet have been captured by the database search. The accepted-paper lists of 10 highly relevant conferences, workshops, and tracks held in 2026 including AIware, ICSE, MSR, CAIN, RAIE, RAISE, AGENT, SEIP, LLMTrust, and RSE were manually screened for potentially relevant studies. The full names and details of these venues are provided in Table~\ref{tab:manual_venues}. This search identified 11 candidate studies, which were subsequently assessed through full-text screening. Seven studies were excluded, while four met the inclusion criteria and were added to the corpus. The final corpus consists of 44 primary studies: 29 identified through the main database search, 11 through snowballing, and four through the manual venue search. To help for traceability throughout the synthesis, each of the 44 selected primary studies was assigned a unique identifier (P01–P44). These identifiers are presented in the primary studies table \ref{tab:primary_studies} and are used throughout the results and discussion to reference individual approaches and support cross-study comparisons.

\begin{table*}[!t]
\centering
\caption{Research venues included in the manual search.}
\label{tab:manual_venues}
\begin{tabular}{p{1.8cm} p{9.0cm} p{2.5cm} p{1.0cm}}
\toprule
\textbf{Acronym} & \textbf{Full Name} & \textbf{Type} & \textbf{Year} \\
\midrule
AIware 
& ACM International Conference on AI-Powered Software 
& Conference 
& 2026 \\

ICSE 
& International Conference on Software Engineering 
& Conference 
& 2026 \\

MSR 
& International Conference on Mining Software Repositories 
& Conference 
& 2026 \\

CAIN 
& International Conference on AI Engineering -- Software Engineering for AI 
& Conference 
& 2026 \\

RAIE 
& International Workshop on Responsible AI Engineering 
& Workshop 
& 2026 \\

RAISE 
& Workshop on Requirements Engineering for AI-Powered Software 
& Workshop 
& 2026 \\

AGENT 
& International Workshop on Agentic Engineering 
& Workshop 
& 2026 \\

SEIP 
& Software Engineering in Practice 
& ICSE Track 
& 2026 \\

LLMTrust 
& International Workshop on Software Engineering for and with Trustworthy LLMs 
& Workshop 
& 2026 \\

RSE 
& Responsible Software Engineering 
& Workshop 
& 2026 \\
\bottomrule
\end{tabular}
\end{table*}



\begin{table*}[!t]
\centering
\caption{Primary studies included in the Systematic Literature Review.}
\label{tab:primary_studies}
\small

\begin{tabularx}{\textwidth}{p{0.06\textwidth} X p{0.18\textwidth} p{0.09\textwidth}}
\toprule
\textbf{Study ID} & \textbf{Paper Title} &
\textbf{Reference} & \textbf{Citation} \\
\midrule
P01 & ML-On-Rails: Safeguarding Machine Learning Models in Software Systems -- A Case Study & Abdelkader et al. (2024) & \cite{ML_on_rails_2024} \\ \addlinespace
P02 & Data-Centric AI Governance for Responsible Organizational Value: Evidence from a European Public Administration & Aguilar et al. (2026)  & \cite{data_centric_AI_governance_2026} \\ \addlinespace
P03 & Inferring Data Preconditions from Deep Learning Models for Trustworthy Prediction in Deployment & Ahmed et al. (2024)  & \cite{deepinfer_2024} \\ \addlinespace
P04 & A Practical Conceptual AI-Driven Framework for Cybersecurity Threat Detection and Secure Deployment & Awajan et al. (2026)  & \cite{ai_cybersecurity_threats_2026} \\ \addlinespace
P05 & Towards MLCertOps in Aviation & Ayyildiz et al. (2025) & \cite{MLCertOps_2020} \\ \addlinespace
P06 & E-FAIR-DB: Functional Dependencies to Discover Data Bias and Enhance Data Equity& Azzalini et al. (2022)  & \cite{EFairDB_2022} \\ \addlinespace
P07 & A Certification Scheme for Large Language Models-Based Applications & Bena et al. (2026)  & \cite{LLM_certification_2026} \\ \addlinespace
P08 & Aspirations and Practice of ML Model Documentation: Moving the Needle with Nudging and Traceability & Bhat et al. (2023)  & \cite{DocML_2023} \\ \addlinespace
P09 & Ergo, SMIRK is safe: a safety case for a machine learning component in a pedestrian automatic emergency brake system & Borg et al. (2023)  & \cite{Ergo_smirk_2023} \\ \addlinespace
P10 & ACR’s Assess-AI: A Registry for Real-World Performance Monitoring of Clinical Imaging Artificial Intelligence & Coombs et al. (2026)  & \cite{assessAI_2026} \\ \addlinespace
P11 & From Policy to Pipeline: A Governance Framework for AI Development and Operations Pipelines & Butt et al. (2026)  & \cite{policy_to_pipeline_2026} \\ \addlinespace
P12 & Discovering and Validating AI Errors With Crowdsourced Failure Reports & Cabrera et al. (2021)  & \cite{deblinder_2021} \\ \addlinespace
P13 & Efficient and Scalable Covariate Drift Detection in Machine Learning Systems with Serverless Computing & Céspedes Sisniega et al. (2024)  & \cite{serverless_covariate_drift_2024} \\ \addlinespace
P14 & An Observability Framework for Detecting Bias and Drift in AI-Based Student Early Alert Systems & Dedu et al. (2025)  & \cite{observability_student_bias_drift_2025} \\ \addlinespace
P15 & An Adaptive Governance-Centric MLOps Framework for Risk-Tiered Control and Continuous Assurance of Responsible AI in High-Stakes Domains & Eraganeni (2026)  & \cite{governance_risk_control_2026} \\ \addlinespace
P16 & Logging requirement for continuous auditing of responsible machine learning-based applications & Foalem et al. (2025)  & \cite{logging_requirements_2025} \\ \addlinespace
P17 & XAI-Compliance-by-Design: A Modular Framework for GDPR- and AI Act-Aligned Decision Transparency in High-Risk AI Systems & Goncalves \& Correia (2026)  & \cite{xai_compliance_2026} \\ \addlinespace
P18 & A Generative Agentic AI Framework for Analysis and Debiasing of Datasets and Machine Learning Models & Gunaweera et al. (2025)  & \cite{Gen_Agent_debias_2025} \\ \addlinespace
P19 & Blockchain-Based Provenance for Artificial Intelligence Lifecycle Traceability & Gunasekara et al. (2026) & \cite{blockchain_provenance_2026} \\ \addlinespace
P20 & Towards Trustworthy AI Engineering: A Case Study on Integrating an AI Audit Catalog into MLOps Processes & Helmer et al. (2024)  & \cite{audit_catolog_case_study_2024} \\ \addlinespace
P21 & AI Data Readiness Inspector (AIDRIN) for Quantitative Assessment of Data Readiness for AI & Hiniduma et al. (2024)  & \cite{AIDRIN_2024} \\ \addlinespace
P22 & The Role of Explainability in Assuring Safety of Machine Learning in Healthcare & Jia et al. (2022)  & \cite{role_of_XAI_healthcare_2022} \\ \addlinespace
P23 & GovernAI: Policy-Driven Model Governance for Dynamic and Multi-Tenant AI Systems & Jonnalagadda et al. (2025)  & \cite{governai_2025} \\ \addlinespace
P24 & AI Model Passport: Data and System Traceability Framework for Transparent AI in Health & Kalokyri et al. (2025)  & \cite{ai_model_passport_2025} \\ \addlinespace
P25 & Monitoring and Explainability of Models in Production & Klaise et al. (2020)  & \cite{monitoring_xai_2020} \\ \addlinespace
P26 & Explaining Shifts in Machine Learning Systems with Causal Maps
 & Leest et al. (2026)  & \cite{tracing_shift_2026} \\ \addlinespace
P27 & Machine Learning Observability in Practice & Leest et al. (2026)  & \cite{monitoring_practices_2026} \\ \addlinespace
\bottomrule
\end{tabularx}
\end{table*}
\begin{table*}[!t]
\centering
\small
\begin{tabularx}{\textwidth}{p{0.06\textwidth} X p{0.18\textwidth} p{0.09\textwidth}}
\toprule
\multicolumn{4}{c}{TABLE III (Continued)} \\
\midrule
\textbf{Study ID} & \textbf{Paper Title} &
\textbf{Reference} & \textbf{Citation} \\
\midrule
P28 & Augur: A Step Towards Realistic Drift Detection in Production ML Systems & Lewis et al. (2023)  & \cite{augur_2023} \\ \addlinespace
P29 & Toward Trustworthy AI: Blockchain-Based Architecture Design for Accountability and Fairness of Federated Learning Systems & Lo et al. (2023)  & \cite{toward_TAI_blockchain_2023} \\ \addlinespace
P30 & Reference Architecture for the Design and Implementation of AI Systems in Manufacturing in Conformity to ISO/IEC 42001 & Mateo-Casalí et al. (2026)  & \cite{ref_architecture_ISO_2026} \\ \addlinespace
P31 & A Semantic Framework to Support AI System Accountability and Audit & Naja et al. (2021) & \cite{semantic_AI_System_accountability_2021} \\ \addlinespace
P32 & Compliance-as-Code for AI-Driven Identity Systems: Clause-to-Control Traceability and Machine-Readable Evidence & Nweke \& Yeng (2026)  & \cite{compliance_as_code_2026} \\ \addlinespace
P33 & An Empirical Study of Policy as Code: Adoption, Purpose, and Maintenance & Opdebeeck et al. (2026)  & \cite{policy_as_code_mining_2026} \\ \addlinespace
P34 & Data Cards: Purposeful and Transparent Dataset Documentation for Responsible AI & Pushkarna et al. (2022)  & \cite{data_cards_2022} \\ \addlinespace
P35 & Hybrid MLOps Framework for Automated Lifecycle Management of Adaptive Phishing Detection Models & Reda et al. (2025)  & \cite{hybridMLOps_2025} \\ \addlinespace
P36 & An Applied Systems Innovation Integrating Blockchain and Deterministic Machine Learning for Trustworthy Artificial Intelligence & Shuitenov et al. (2026)  & \cite{HyCoss_2026} \\ \addlinespace
P37 & LIFEDATA: A Framework for Traceable Active Learning Projects & Stieler et al. (2023)  & \cite{lifedata_2023} \\ \addlinespace
P38 & The SPATIAL Architecture: Design and Development Experiences from Gauging and Monitoring the AI Inference Capabilities of Modern Applications & Ottun et al. (2024)  & \cite{spatial_architecture_2024} \\ \addlinespace
P39 & The Use of eXplainable Artificial Intelligence and Machine Learning Operation Principles to Support the Continuous Development of Machine Learning-Based Solutions in Fault Detection and Identification & Tran et al. (2024)  & \cite{xai_mlops_fault_detection_2024} \\ \addlinespace
P40 & Governance and Auditability of AI-Driven Retail Decision Pipelines in Cloud-Native Architectures & Veluchamy (2026)  & \cite{AI_driven_retail_decision_2026} \\ \addlinespace
P41 & How Are MLOps Frameworks Used in Open Source Projects? An Empirical Characterization & Zampetti et al. (2026)  & \cite{opensource_mlops_2026} \\ \addlinespace
P42 & Continuous Development and Safety Assurance for ML in Railway & Zeller et al. (2024)  & \cite{CD_safety_railway_2024} \\ \addlinespace
P43 & Secure AI-SDLC for Critical Infrastructure: Operationalizing the NIST AI RMF with Evidence-Driven Controls & Sudarsan et al. (2025)  & \cite{AI_SDLC_2025} \\ \addlinespace
P44 & MOSAIC: Model-Based Safety Analysis Framework for AI-Enabled Cyber-Physical Systems & Xie et al. (2026)  & \cite{MOSAIC_2026} \\
\bottomrule
\bottomrule
\end{tabularx}
\end{table*}


\begin{table*}[t]
\centering
\caption{Data extraction fields used for synthesis.}
\label{tab:extraction}
\begin{tabularx}{\textwidth}{p{0.23\textwidth} X}
\toprule
\textbf{Extraction field} & \textbf{Purpose} \\
\midrule
Study metadata & Title, authors, year, venue, DOI, and source database. \\
Problem addressed & Identifies the operational Responsible AI problem targeted by the study. \\
Responsible AI concerns & Identifies fairness, accountability, transparency, explainability, privacy, security, safety, robustness, compliance, human oversight, and sustainability. \cite{SR_Ethics_tools_2024} \\
ML lifecycle stage & Maps the mechanism to data, development, validation, deployment, monitoring, retraining, governance/audit, or incident response. \cite{MLR_MLOps_practices}\\
Operational mechanism & Describes the concrete mechanism, such as runtime monitor, provenance framework, audit artifact, policy gate, guardrail, or evidence bundle. \cite{RAI_pattern_catalogue_2024} \\
Evaluation approach & Explains whether the paper provides prototype demonstration, benchmark experiment, case study, deployment, user study, expert validation, simulation, or no evaluation. \\
Evidence artifacts & Identifies logs, alerts, reports, model cards, lineage records, certificates, assurance cases, compliance evidence, or governance records. \\
Limitations and future work & Supports gap analysis and future research direction. \\
\bottomrule
\end{tabularx}
\end{table*}

\subsection{Synthesis and Data Extraction}
\label{sec:synthesis-extraction}
Before synthesizing the final corpus of 44 studies by research question, we first mapped the corpus descriptively to understand when the literature emerged, where it was published, which domains it addressed, what Responsible AI concerns were covered, and how mature the reported evaluations were.

The synthesis combined descriptive mapping with thematic analysis. Descriptive mapping was used to characterize the corpus by publication year, venue type, application domain, Responsible AI concern, ML lifecycle stage, operational mechanism, and evaluation maturity. This mapping provides an overview of where the literature is concentrated and where evidence remains limited \cite{sms_2015}.

Thematic analysis was then used to group studies according to the primary mechanism through which Responsible AI is operationalized in ML production \cite{thematic_analysis_2011}. Rather than organizing the synthesis only by high-level Responsible AI principles, we used mechanism-based themes because the review focuses on how Responsible AI is implemented in practice. The resulting themes were: monitoring and runtime assurance; audit and compliance; Responsible MLOps lifecycle integration; and traceability and provenance. A smaller set of studies focused on pre-model data and bias safeguards and was analyzed as a supporting theme.

The Responsible AI concerns, including fairness, safety, security, explainability, privacy, sustainability, and human oversight\cite{SR_Ethics_tools_2024}, were analyzed as secondary as these concerns appeared across multiple mechanism categories. For example, fairness was operationalized through data documentation ~\cite{data_cards_2022}, bias auditing~\cite{Gen_Agent_debias_2025}, monitoring, and governance workflows ~\cite{governai_2025,observability_student_bias_drift_2025}, while explainability appeared in safety assurance~\cite{role_of_XAI_healthcare_2022}, compliance, and runtime debugging studies~\cite{monitoring_xai_2020,xai_mlops_fault_detection_2024}. Therefore, treating these concerns as secondary dimensions preserved the relationships between Responsible AI values and the operational mechanisms used to implement them.

\section{Results}
\subsection{Data Synthesis Overview}
\label{results}
After completing the data extraction, the results were summarized based on publication year, theme coverage, ML lifecycle-stage coverage, and evaluation maturity. These summaries were used to identify dominant patterns in the literature and to support the gap analysis.

 \begin{figure}[t]
    \centering
    \includegraphics[width=\linewidth]{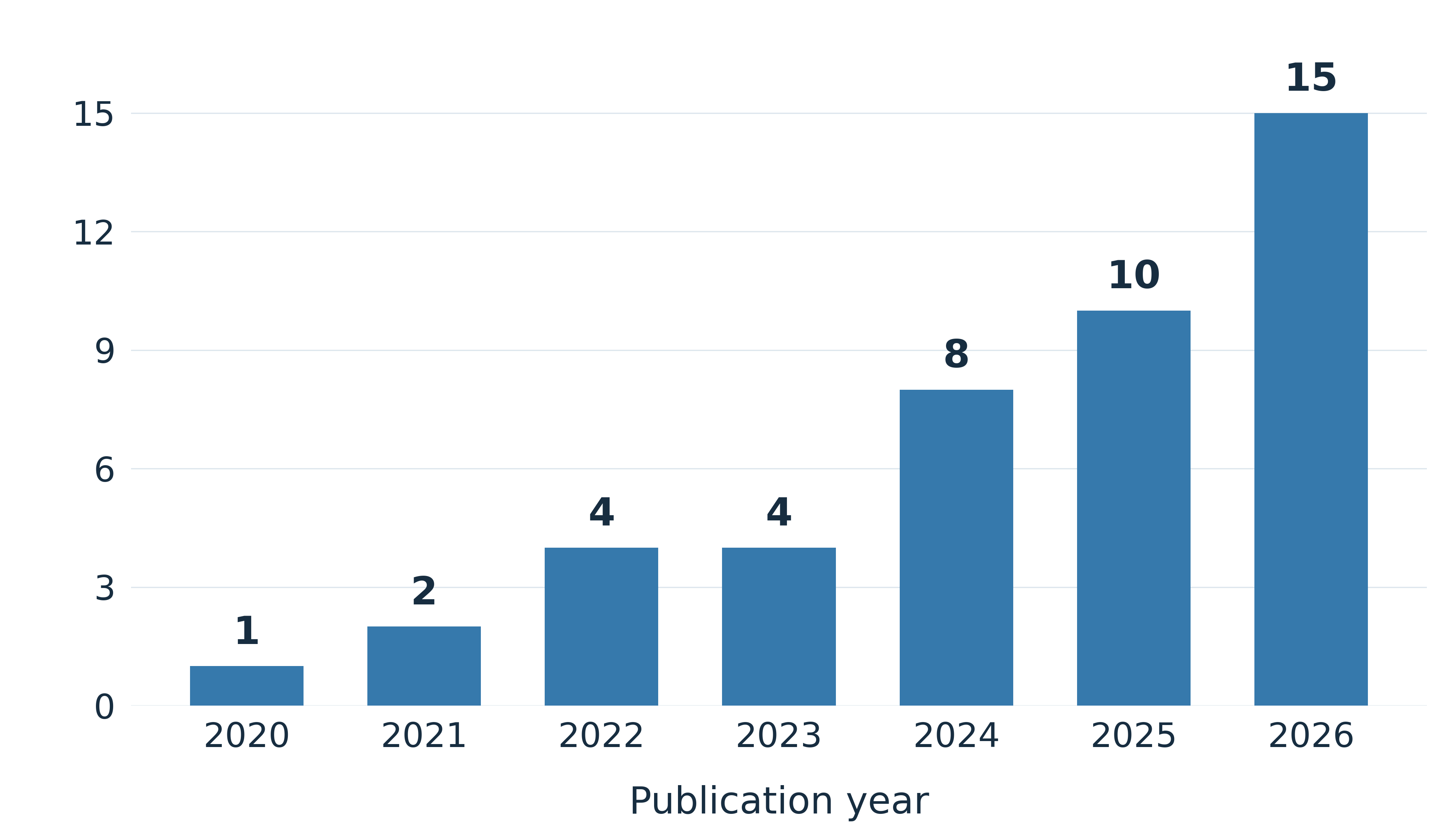}
    \caption{The publication timeline of the research studies.}
    \label{fig:timeline}
\end{figure}

Figure~\ref{fig:timeline} shows that the literature is recent and has grown substantially after 2023. This trend reflects the increasing attention to operational Responsible AI, especially as MLOps, AI governance, and regulatory compliance have become more closely connected. The corpus includes both conference and journal publications, indicating that the topic is getting more attention across software engineering, AI engineering, information systems, healthcare AI, and safety-critical systems venues.

 \begin{figure}[t]
    \centering
    \includegraphics[width=\linewidth]{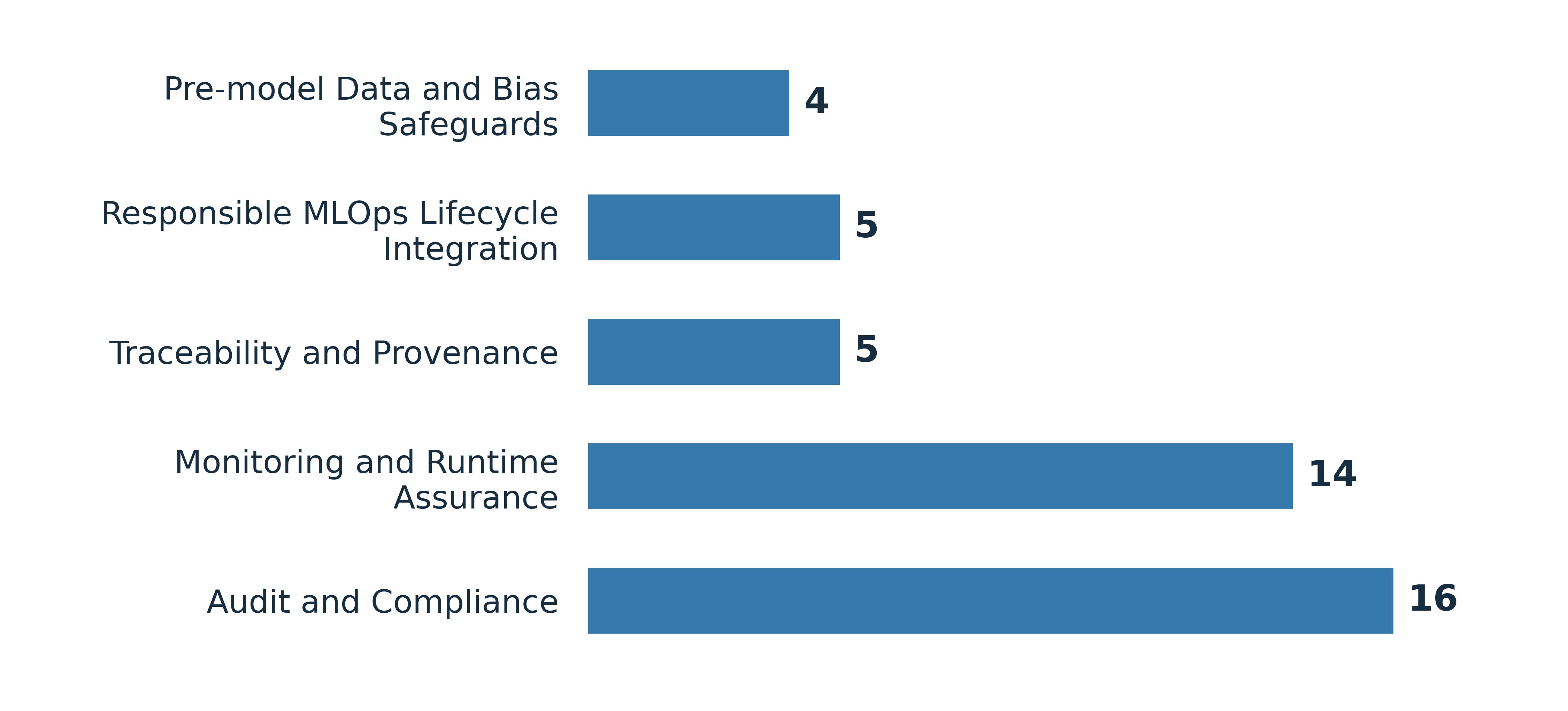}
    \caption{The distribution of mechanism-based themes across studies}
    \label{fig:themes}
\end{figure}

Figure~\ref{fig:themes} summarizes the mechanism-based themes used for synthesis. The results converged into five primary themes. \emph{Monitoring and runtime assurance} refers to mechanisms applied after deployment, including drift detection, outlier detection, runtime explainability, observability, and feedback loops for retraining or human review~\cite{monitoring_xai_2020, augur_2023}. \emph{Audit and compliance} captures studies that convert governance, regulatory, ethical, or safety expectations into audit procedures, certification evidence, compliance checks, assurance cases, or machine-readable controls~\cite{compliance_as_code_2026, MLCertOps_2020}. \emph{Responsible MLOps lifecycle integration} includes studies that embed Responsible AI into ML pipelines through gates, approval workflows, retraining processes, deployment controls, and ML lifecycle governance~\cite{ML_on_rails_2024, governai_2025}. \emph{Traceability and provenance} covers dataset and model documentation, lineage tracking, metadata capture, audit logs, evidence artifacts, and reproducibility mechanisms that support accountability~\cite{data_cards_2022,ai_model_passport_2025,HyCoss_2026}. A smaller group of \emph{pre-model safeguards} focus on data readiness, bias detection, dataset repair, or documentation that strengthens production governance before model training or deployment~\cite{AIDRIN_2024, Gen_Agent_debias_2025}.

 \begin{figure}[t]
    \centering
    \includegraphics[width=\linewidth]{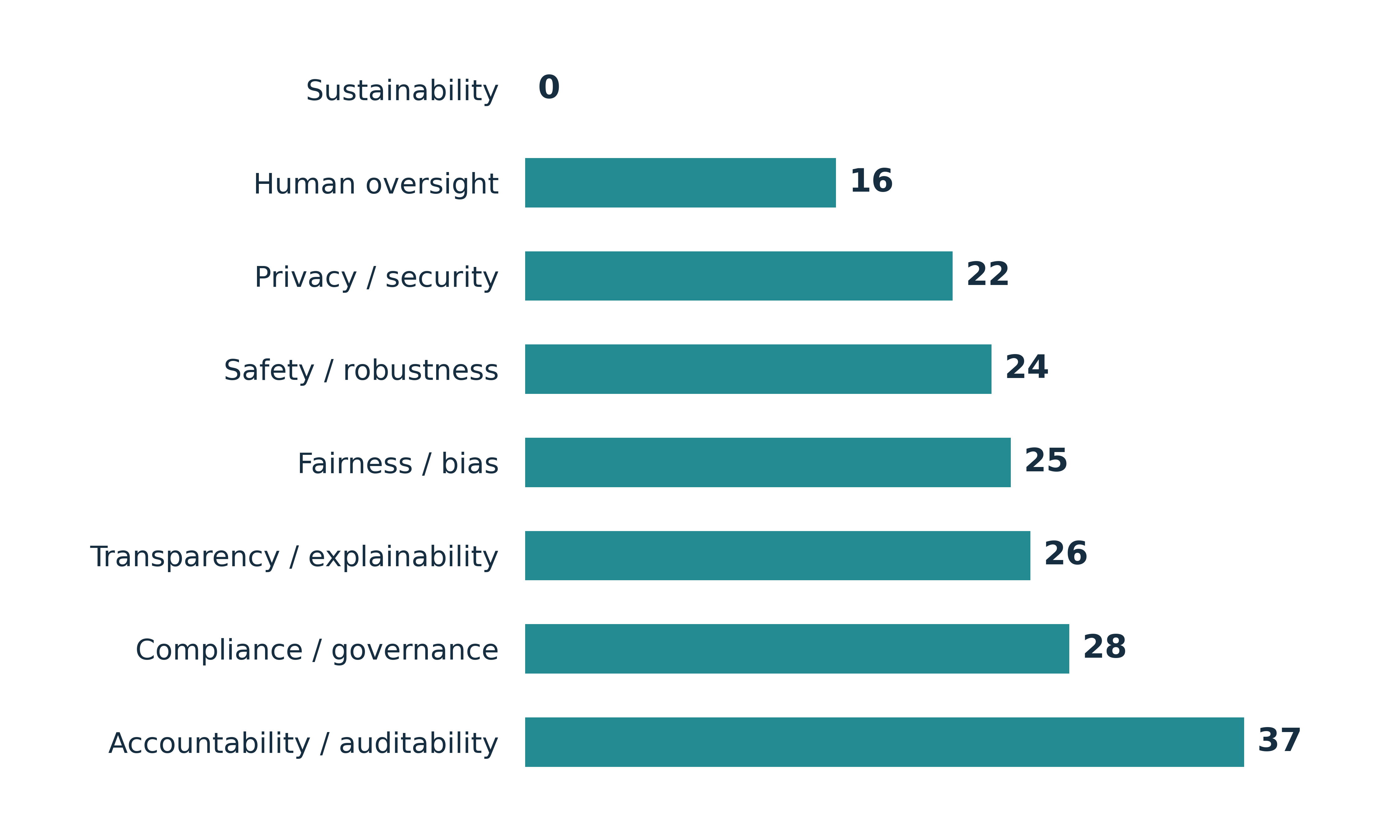}
    \caption{The distribution of Responsible AI principles addressed across studies}
    \label{fig:rai-concerns}
\end{figure}

Figure~\ref{fig:rai-concerns} shows that most studies address multiple Responsible AI concerns simultaneously. As discussed in Section~\ref{sec:synthesis-extraction}, accountability, auditability, compliance, transparency, fairness, safety, and privacy are treated as interconnected concerns rather than separate topics. The strongest concentration is around accountability and auditability, followed by compliance and governance. This indicates that the operational Responsible AI literature is increasingly concerned with making AI systems inspectable, reviewable, and defensible within organizational or regulatory workflows, rather than only improving model behavior in isolation.

Transparency and explainability, fairness and bias, safety and robustness, and privacy and security also appear frequently, but they are usually embedded within broader operational mechanisms. This suggests a shift from treating Responsible AI values as abstract principles toward implementing them as pipeline controls, evidence artifacts, monitoring signals, and assurance mechanisms.

The relatively lower coverage of human oversight suggests that although many studies acknowledge the need for human review, fewer provide concrete mechanisms for assigning responsibility, integrating human decisions into production workflows, or evaluating human-AI collaboration in operational settings. Sustainability is absent from the coded corpus, which indicates that environmental concerns are not yet strongly integrated into the operational Responsible AI and MLOps literature represented in this review. Overall, the trend suggests that current work is most mature where Responsible AI can be translated into auditable artifacts and technical controls, while softer organizational concerns such as human responsibility and sustainability remain less operationalized.

 \begin{figure}[t]
    \centering
    \includegraphics[width=\linewidth]{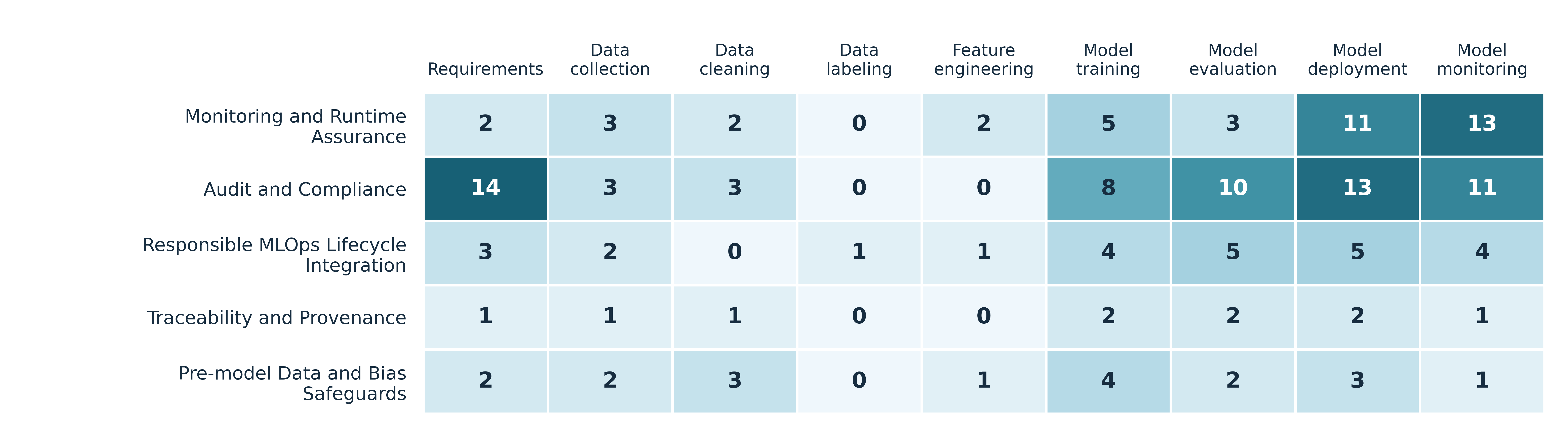}
    \caption{The distribution of studies across the MLOps lifecycle stages}
    \label{fig:lifecycle-heatmap}
\end{figure}

Figure~\ref{fig:lifecycle-heatmap} maps the themes against ML lifecycle stages based on the software engineering for machine learning workflow described by Amershi et al.~\cite{SE_in_ML_2019}, which includes model requirements, data collection, data cleaning, data labeling, feature engineering, model training, model evaluation, model deployment, and model monitoring. Governance, audit, and assurance mechanisms were not treated as separate ML lifecycle stages; instead, they were mapped to the stage where they operationally included.

The heatmap shows that the corpus is concentrated around deployment and monitoring, followed by model training, evaluation, and requirements. This pattern reflects the review focus on operational Responsible AI where many studies are concerned with how models are governed, released, observed, audited, or corrected after development. Earlier ML lifecycle stages such as data labeling and feature engineering receive comparatively less explicit attention, except in data-centric and documentation-oriented studies. This imbalance suggests that the literature has moved toward post-development controls, while earlier data-production activities remain less systematically connected to Responsible AI governance in MLOps pipelines.

 \begin{figure}[t]
    \centering
    \includegraphics[width=\linewidth]{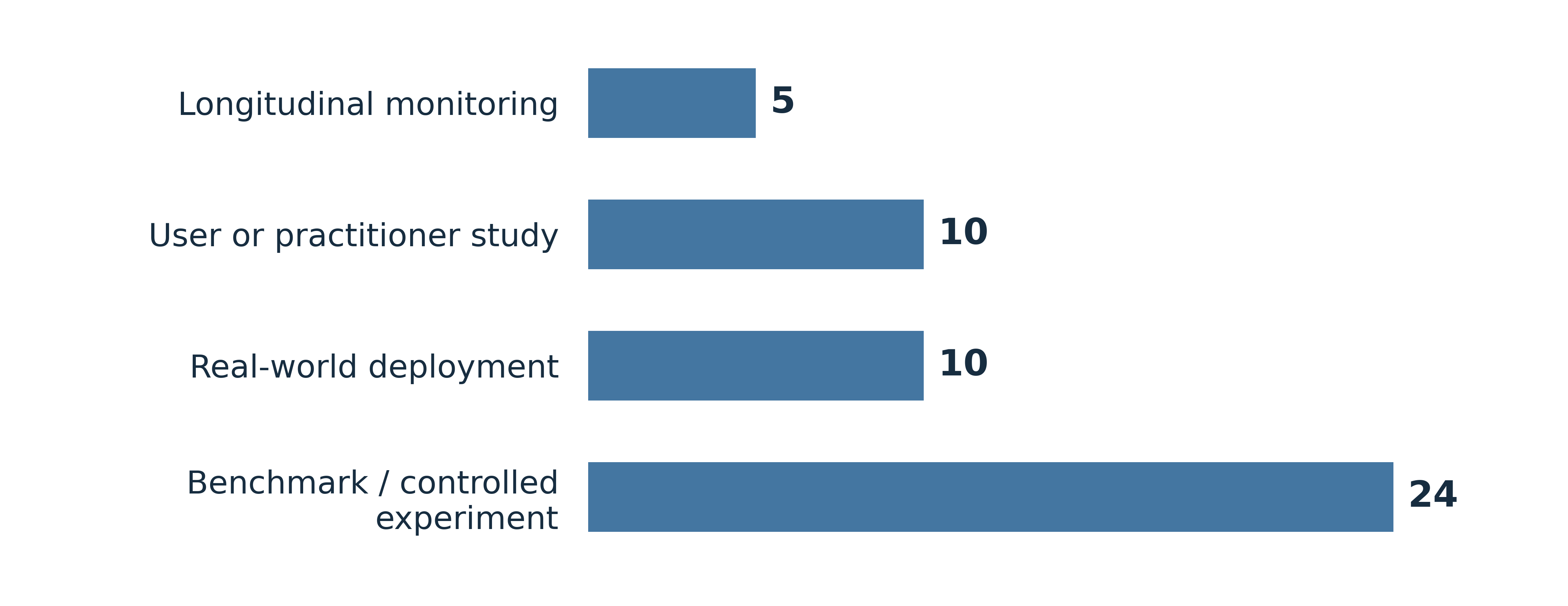}
    \caption{Distribution of primary studies by evaluation maturity.}
    \label{fig:evaluation}
\end{figure}

Figure~\ref{fig:evaluation} summarizes the evaluation maturity of the corpus. As the operationalisation of Responsible AI is a developing research area, it is important to analyse the evaluation maturity of the studies to understand the research directions. The studies range from conceptual frameworks and prototypes to benchmark experiments, case studies, practitioner studies, and deployment-oriented evaluations. Although several papers provide strong technical prototypes or controlled experiments, fewer demonstrate long-term use in real production environments. This pattern motivated the discussion of adoption barriers and evidence gaps in RQ3.

Overall, the descriptive synthesis shows that the corpus is technically diverse but unevenly studied. The literature has moved beyond abstract Responsible AI principles toward concrete mechanisms such as drift detection, bias monitoring, compliance-as-code, evidence bundles, and ML lifecycle gates. However, the maturity and operational validation of these mechanisms vary across themes and domains.

To examine these findings in depth, the next section moves from descriptive mapping to a discussion organized around the three RQs. We analyze how the selected studies address monitoring and lineage (RQ1), implement pipeline-level governance and deployment controls (RQ2), and deal with socio-technical challenges in real-world development settings (RQ3).

\section{Discussion}
\label{discussion}
In this section, we analyze the 44 primary studies to directly answer our three RQs.
\subsection{RQ1: Runtime Observability, Statistical Drift Monitoring, and Lineage Mechanisms}
\label{RQ1}
Continuous assurance in MLOps depends on the ability to observe system behavior and maintain trustworthy records of how data and models evolve over time. RQ1 therefore focuses on the monitoring and lineage mechanisms that support these capabilities \cite{ai_model_passport_2025, monitoring_xai_2020, blockchain_provenance_2026}. It examines the technical architectures, statistical methods, and cryptographic or semantic approaches used to monitor incoming data and model behavior, detect distribution shifts or performance degradation, and preserve reliable records of datasets, model versions, pipeline executions, and deployment decisions. Together, these mechanisms provide the evidence required to investigate failures, reconstruct system behavior, and support auditing and accountability. 

\subsubsection{\textbf{Asynchronous Runtime Telemetry Architectures}}\mbox{}\\ 
A key challenge in production observability is assessing model health without adding significant latency to synchronous inference paths. Traditional software monitoring primarily focuses on system-level metrics such as server load, memory usage, request frequency, and response latency \cite{hybridMLOps_2025, ML_on_rails_2024}. ML systems require additional monitoring of feature distributions, correlations, prediction outputs, and changes in model behavior \cite{monitoring_xai_2020}. Performing computationally intensive statistical analyses directly within prediction handlers can therefore increase latency and reduce serving throughput \cite{serverless_covariate_drift_2024, ai_cybersecurity_threats_2026}. To address this limitation, several studies adopt \textbf{decoupled, event-driven telemetry architectures}, where inference and monitoring are executed independently \cite{hybridMLOps_2025, ref_architecture_ISO_2026}. 

Seldon Core and Alibi Detect implement the \textit{Black-Box Explainer and Outlier Pattern}, in which inference payloads are asynchronously forwarded through Knative Eventing to separate monitoring components \cite{monitoring_xai_2020}. Céspedes Sisniega et al. \cite{serverless_covariate_drift_2024} extend this approach by combining serverless infrastructure with Autoencoders (AEs) that reduce high-dimensional inputs to lower-dimensional representations. Serverless \textit{Frouros} functions can then compute drift measures such as the Kolmogorov--Smirnov (KS) statistic, Hellinger distance, and Maximum Mean Discrepancy (MMD) outside the latency-sensitive inference path. A related approach is presented by Ottun et al. \cite{spatial_architecture_2024} through \textbf{SPATIAL}, which instruments ML applications with modular, out-of-band monitoring microservices. These components observe properties such as adversarial robustness, explanation stability, and distribution drift and expose the resulting indicators through an interactive dashboard. This illustrates a broader pattern of separating runtime assurance mechanisms from the prediction service while still providing operators with visibility into model behavior. 

Complementing runtime monitoring architectures, Lewis et al. \cite{augur_2023} examine the pre-deployment problem of determining which drift measures are most suitable for a particular application. Their \textbf{Augur} framework introduces controlled prevalence and covariate drift during development and evaluates multiple statistical measures against resulting changes in model performance. Their findings indicate that different measures capture different characteristics of degradation: KL Divergence can provide an early indication of change under time-series uncertainty, whereas Energy Distance can provide more steadily accumulating evidence of sustained degradation. These results suggest that drift monitoring should be selected and calibrated according to the deployment context rather than relying on a single universal metric. Similar patterns appear across application domains. In cybersecurity and network intrusion detection, Reda et al. \cite{hybridMLOps_2025} use a microservices-based monitoring architecture that combines statistical drift measures, including the Population Stability Index (PSI), KL Divergence, and KS tests, with changes in SHAP feature attributions. Their implementation reports drift detection within 18 seconds while maintaining approximately 50\,ms inference latency at 2,300 requests per second. In educational early-alert systems, Prometheus-, Loki-, and Grafana-based telemetry is used to monitor concept drift and demographic bias \cite{observability_student_bias_drift_2025}. Similarly, Tran et al. \cite{xai_mlops_fault_detection_2024} combine LSTM-based fault detection with SHAP/LIME telemetry and Plan-Do-Check-Act (PDCA) feedback loops to distinguish sensor noise from broader signal drift in industrial systems. 

Overall, the reviewed studies identify two complementary requirements for production ML observability. First, monitoring should be separated from latency-sensitive inference through asynchronous or event-driven architectures \cite{monitoring_xai_2020, serverless_covariate_drift_2024}. Second, the statistical measures used for monitoring should be selected and calibrated according to the characteristics of the deployment environment \cite{augur_2023}. Together, these approaches enable continuous observation of model behavior without imposing excessive overhead on production inference.

\subsubsection{\textbf{Diagnostic Shift Attribution}}\mbox{}\\
Statistical drift tests can indicate that a distribution has changed, but they often provide limited information about the cause of that change. This makes it difficult to determine whether an observed shift results from an internal pipeline issue, such as a parsing error, or from an external change, such as seasonal variation. Poorly explained alerts can also contribute to alert fatigue. To address this limitation, Leest et al. \cite{tracing_shift_2026} introduce \textbf{Causal System Maps} based on Structural Causal Models (SCMs). Their approach represents the MLOps environment through causally linked views of the \textit{Environment}, \textit{ML System}, and \textit{Subsystems}, with variables categorized as Data ($D$), Modulator ($M$), or Random ($R$) nodes. When drift is detected, Shapley-based change attribution is used to trace the anomaly through the causal structure and identify its likely source. This extends conventional drift detection from simply identifying that a change has occurred to explaining where it originated. Such diagnostic attribution can support more targeted debugging, distinguish internal faults from external environmental changes, and reduce unnecessary responses such as model retraining.

\subsubsection{\textbf{Verifiable Data and Model Lineage}}\mbox{}\\
The literature on lineage tracking shows a shift from conventional, mutable logs toward \textbf{cryptographically verifiable and content-addressed provenance}. In production MLOps, reliable lineage requires demonstrating that the datasets, preprocessing code, configurations, and model artifacts approved during validation correspond to those ultimately deployed in production. Such traceability supports auditability and helps detect unauthorized or undocumented changes. Frameworks such as HyCoSS/HyCoSS++ \cite{HyCoss_2026} provide this assurance through a \textbf{hash-based binding approach}. Dataset partitions, preprocessing scripts, hyperparameter configurations, and trained model weights are represented within a content-addressed Merkle Directed Acyclic Graph (Merkle-DAG). A cryptographic Merkle root representing these artifacts is then anchored to a distributed ledger or smart contract, creating a verifiable Proof-of-Lineage (PoL). If an underlying dataset, configuration, source file, or model artifact is modified, its digest changes and produces a different Merkle root that no longer matches the ledger-anchored value. Undocumented modifications therefore become detectable.

Lineage mechanisms must also support workflows in which datasets and models evolve continuously. In human-in-the-loop settings, LIFEDATA \cite{lifedata_2023} combines Git and Data Version Control (DVC) to jointly version datasets, annotation activities, annotator information, and active-learning checkpoints. Similarly, in multi-tenant cloud environments, GovernAI \cite{governai_2025} records model ancestry, branching relationships, and execution histories in an append-only governance ledger. Related blockchain-based approaches \cite{blockchain_provenance_2026, toward_TAI_blockchain_2023} combine ledger-based provenance with policy validation and access-control mechanisms to maintain verifiable records of model aggregation and configuration updates. 

Overall, these approaches extend lineage beyond simple version identifiers and mutable logs. By linking ML artifacts through content hashes, version-control systems, and append-only records, they provide tamper-evident evidence of how datasets, configurations, and models evolve throughout the ML lifecycle.

\subsubsection{\textbf{Semantic Provenance and Knowledge Graphs}}\mbox{}\\ 
While cryptographic lineage helps establish artifact integrity, it does not capture the meaning and relationships among ML lifecycle activities. In clinical and other safety-critical domains, the literature therefore introduces \textbf{semantic provenance models} that represent relationships among datasets, processing activities, models, and human decisions in a structured and queryable form. 

Kalokyri et al. (\textbf{AIPassport}) \cite{ai_model_passport_2025} and Naja et al. (\textbf{SAO/RAInS}) \cite{semantic_AI_System_accountability_2021} use semantic provenance models based on standards such as W3C PROV-O and the W3C Machine Learning Schema (MLS). Activities including data curation, de-identification, model training, and approval decisions are represented as connected entities and processes across the ML lifecycle. When serialized as RDF triples and integrated with metadata catalogs such as HealthDCAT-AP and EUCAIM, these provenance graphs can be queried using SPARQL to trace dataset origins, inspect preprocessing and anonymization activities, and identify recorded human approvals. AIPassport further presents this machine-readable provenance through human-readable interfaces, making the same evidence accessible to both automated systems and auditors. 

Together, these studies show that semantic provenance complements cryptographic lineage by adding contextual meaning to ML lifecycle records. Knowledge graphs provide a structured and queryable representation of how data, models, processing activities, and governance decisions are connected, strengthening traceability and supporting automated compliance analysis.

\subsubsection{\textbf{Pre-Training Data Quality and Bias Safeguards}}\mbox{}\\ 
The reviewed studies show that Responsible AI assurance should begin before model training, with safeguards that assess data quality, suitability, fairness, and privacy during data ingestion and preparation. Hiniduma et al. (\textbf{AIDRIN}) \cite{AIDRIN_2024} propose a pre-training data-readiness assessment based on the FAIR principles---Findable, Accessible, Interoperable, and Reusable---together with measures of class imbalance and privacy risk. These include the Likelihood Ratio Imbalance Degree (LRID) for assessing representation imbalance and Markov-model-based risk scores for estimating re-identification risk. By identifying such issues before training, AIDRIN supports early intervention rather than relying solely on post-deployment monitoring. 

In relational database settings, Azzalini et al. (\textbf{E-FAIR-DB}) \cite{EFairDB_2022} use Approximate Conditional Functional Dependencies (ACFDs) to detect and rank potentially discriminatory relationships in tabular data. The approach examines dependencies between protected attributes, such as race or gender, and outcome variables, such as income or loan approval. This allows data engineers to identify problematic patterns and address potential inequities before model training begins. Together, these studies extend continuous assurance upstream into the data preparation stage. Rather than treating bias, privacy, and data quality as issues to be detected only after deployment, they introduce preventive safeguards that reduce downstream fairness, privacy, and reliability risks.
\subsection{RQ2: Codified Governance, Policy-as-Code, and Pipeline-Level Controls} 
\label{RQ2}
Moving AI governance from manual, retrospective documentation toward continuous assurance requires legal, ethical, and safety requirements to be translated into machine-interpretable rules and integrated directly into MLOps pipelines. RQ2 examines how the reviewed studies operationalize requirements from frameworks such as the EU AI Act, GDPR, eIDAS 2.0, ISO/IEC 42001, and NIST AI RMF 1.0 through executable policies, automated deployment gates, and structured safety and compliance evidence. 

\subsubsection{\textbf{Policy-as-Code and Regulatory Control Mapping}}\mbox\\
Regulatory requirements are typically expressed in natural language, creating a gap between high-level obligations and the technical controls that can be enforced within ML pipelines. Several studies address this gap through \textbf{Policy-as-Code (PaC)}, where governance requirements are translated into machine-executable policies and evaluated automatically at different stages of the ML lifecycle. Policy engines such as the Open Policy Agent (OPA), using Rego, are commonly used to implement these controls. 

Nweke et al. \cite{compliance_as_code_2026} propose the \textbf{O-C-K-E model}, which links regulatory obligations to technical controls, executable checks, and generated evidence. In this approach, compliance requirements are evaluated at deployment gates, and model promotion can be blocked when required evidence is missing or cannot be verified. The resulting evidence can also be exported into standardized formats such as OSCAL, providing traceability between regulatory requirements, technical controls, and pipeline outcomes. Similarly, Goncalves and Correia \cite{xai_compliance_2026} introduce a \textbf{Compliance-by-Design Engine (CDE)} that combines technical execution with a parallel compliance flow. Their approach applies evidence-based gates across data, model, explanation, and audit stages, producing machine-readable records that link pipeline decisions to regulatory requirements. 

\textbf{GovernAI} \cite{governai_2025} follows a related pattern by embedding policy checks into model registration, validation, and deployment, allowing violations of requirements such as fairness or out-of-distribution thresholds to trigger automated actions including model quarantine or revocation. Other studies extend policy-based governance to application-specific settings. For example, Veluchamy \cite{AI_driven_retail_decision_2026} incorporates traceability and automated policy objectives into cloud-native decision pipelines for real-time retail applications. 

Across these studies, a common pattern emerges: compliance is increasingly treated as an executable part of the ML lifecycle rather than as a retrospective documentation activity. Regulatory requirements are mapped to technical checks, evaluated at pipeline gates, and linked to machine-readable evidence. This enables continuous assurance by making governance decisions directly enforceable within operational MLOps workflows.

\subsubsection{\textbf{Pipeline-Native Promotion and Risk-Adaptive Controls}}\mbox{}\\

When compliance checks are performed separately from the deployment process, non-compliant or insufficiently verified models may still reach production. To reduce this risk, several studies integrate governance controls directly into CI/CD and MLOps pipelines, making compliance checks part of the model promotion process. Butt et al. \cite{policy_to_pipeline_2026} propose a fail-closed promotion approach in which PaC rules are evaluated at multiple ML lifecycle stages, including data preparation, training, validation, release, and operations. Model promotion is allowed only when the required evidence satisfies the applicable policies, while the resulting artifacts are preserved as signed and traceable compliance evidence. This represents a broader pattern of \textit{evidence-driven promotion gates}, where governance requirements are enforced before a model can advance through the pipeline. 

Other studies introduce \textbf{risk-adaptive gatekeeping}. The study proposed by Eraganeni et al. \cite{governance_risk_control_2026} adjusts governance requirements according to the assessed risk level of a model. Higher-risk deployments are subject to stricter controls, such as fairness evaluation, explainability checks, and human approval, whereas lower-risk systems can follow lighter assurance paths. This reduces the limitations of applying identical governance requirements to systems with different risk profiles. Security-focused approaches follow a similar pipeline-native pattern. Sudarsan et al. \cite{AI_SDLC_2025} and Reda et al. \cite{hybridMLOps_2025} integrate security and risk-management activities into CI/CD workflows, including adversarial robustness testing, threat assessment, and artifact verification before deployment. Together, these studies show that continuous assurance is increasingly implemented through pipeline-native promotion controls. Rather than treating governance as a separate review activity, policies are evaluated during model transitions, with the strictness of the controls determined by the deployment context and risk level. This enables governance decisions to directly influence whether a model is promoted, blocked, or escalated for additional review.

\subsubsection{\textbf{Domain-Specific Continuous Assurance and Safety Cases}}\mbox{}\\ 

In safety-critical domains, Policy-as-Code is often complemented by domain-specific assurance mechanisms that connect high-level safety or compliance goals to concrete technical evidence. These mechanisms include structured safety cases, runtime monitoring, explainability, human review, and continuous recertification. Goal Structuring Notation (GSN) is commonly used to organize assurance arguments and link system-level claims to validation results, operational evidence, and human oversight. Borg et al. \cite{Ergo_smirk_2023} demonstrate this pattern in automotive systems through SMIRK, where a GSN-based safety case is combined with an out-of-distribution (OOD) monitoring mechanism. Inputs detected outside the expected operating distribution can trigger a safety response that prevents potentially unreliable ML outputs from directly influencing system actions. This illustrates how runtime safety controls can be connected to structured assurance arguments rather than being evaluated only before deployment. 

In healthcare, Jia et al. \cite{role_of_XAI_healthcare_2022} incorporate explainability evidence into clinical safety cases. Feature-attribution and counterfactual methods support human assessment of whether model predictions are consistent with clinically meaningful relationships. Explainability therefore serves not only as an interpretability mechanism, but also as evidence within a broader assurance process. A related clinical approach is presented by Coombs et al. \cite{assessAI_2026} through ACR Assess-AI, which supports post-deployment monitoring of imaging AI across multiple healthcare facilities. The system uses de-identified imaging metadata and radiology reports to derive surrogate diagnostic labels and compare them with model predictions. When performance degradation or disagreement is detected, cases can be reviewed by clinical administrators and the outcomes fed back into the monitoring process. This demonstrates a \textit{closed-loop assurance} model in which automated monitoring is combined with domain-expert review and follow-up investigation. 

Other studies extend assurance beyond one-time certification. Zeller et al. \cite{CD_safety_railway_2024} and Ayyildiz et al. \cite{MLCertOps_2020} connect MLOps pipelines with certification workflows so that operational monitoring can trigger updates to assurance evidence and renewed safety evaluation when system conditions change. This reflects a broader pattern of \textit{continuous recertification}, where safety arguments and supporting evidence evolve alongside the deployed system. Runtime guardrails provide an additional layer of protection in autonomous and generative systems. Xie et al. \cite{MOSAIC_2026} and Abdelkader et al. \cite{ML_on_rails_2024} use formal verification and runtime control mechanisms to constrain unsafe system behavior during operation. These approaches show how explicit safety properties can be enforced alongside continuous monitoring rather than assessed only at development time. 

Overall, the reviewed studies show that continuous assurance in safety-critical domains relies on a combination of structured safety arguments, runtime monitoring, explainability, human review, formal verification, and recertification mechanisms. Across these approaches, the common goal is to keep assurance evidence connected to actual system behavior throughout the ML lifecycle, enabling safety and compliance decisions to be updated as models, data, and operating conditions change.

\subsection{RQ3: Operational Realities and Socio-Technical Friction} 
\label{RQ3}
Although the literature proposes increasingly sophisticated mechanisms for continuous monitoring, Responsible AI (RAI), and policy-driven governance, their adoption in operational MLOps remains uneven. RQ3 examines the practical, organizational, and technical barriers that limit the implementation of these mechanisms in real-world workflows. The reviewed studies reveal a recurring gap between the capabilities proposed in research and the practices sustained in production. This gap is influenced not only by technical limitations, but also by integration costs, stakeholder expertise, organizational incentives, human factors, and the increasing non-determinism of generative and agentic AI systems.

\subsubsection{\textbf{The Aspiration--Practice Gap in Operational MLOps}}\mbox{}\

The reviewed studies consistently reveal a significant gap between Responsible AI mechanisms proposed in the literature and those adopted in everyday ML development. We characterize this disconnect as the \textbf{Aspiration--Practice Gap}. Although governance frameworks increasingly emphasize continuous monitoring, auditability, fairness, explainability, privacy, and evidence generation, these capabilities remain weakly represented in operational workflows.

One indication of this gap is the limited logging of Responsible AI information. Foalem et al. \cite{logging_requirements_2025}, analyzing 85 collaborative open-source ML repositories that use libraries such as Fairlearn, AIF360, Opacus, Captum, SHAP, and PySyft, find that developers predominantly log conventional technical information such as training metrics, execution parameters, and system performance. In contrast, fairness measures, privacy parameters, subgroup performance, and explanation-related evidence are rarely preserved in runtime logs. Responsible AI telemetry therefore remains outside traditional software logging practice, creating an \textbf{auditability gap} in which deployed systems may lack the evidence necessary to reconstruct or assess their ethical behaviour after deployment.

This gap is reinforced by practical adoption barriers. Zampetti et al. \cite{opensource_mlops_2026}, based on an analysis of 53,925 open-source repositories, report substantially greater adoption of lightweight experiment-tracking tools than workflow orchestration, post-deployment validation, and drift-monitoring mechanisms. Tools that provide immediate development value are more readily adopted, whereas mechanisms requiring pipeline restructuring, additional dependencies, or substantial refactoring face greater resistance. Infrastructure incompatibility and the limited availability of lightweight auditing tools further increase this burden. Consequently, Responsible AI controls are often treated as optional extensions rather than normal components of the MLOps lifecycle.

Organizational factors further contribute to this problem. Different stakeholders, including software engineers, ML engineers, domain experts, business teams, and governance personnel, often possess different levels of technical and Responsible AI expertise. This makes it difficult to determine who should define, implement, validate, and maintain governance requirements. In addition, organizations may lack clear incentives, ownership structures, or top-down commitment for Responsible AI activities that do not provide an immediate functional benefit. These factors suggest that the implementation gap is not caused by the absence of technical mechanisms alone, but by the additional effort required to integrate them into existing development practices.

A further limitation is the lack of interactive development environments that actively encourage or enforce Responsible AI practices during implementation. Most governance activities remain separate from developers' normal workflows, increasing the likelihood that fairness checks, documentation, traceability, or audit evidence are created retrospectively or omitted entirely. Overall, the literature indicates a need for Responsible AI mechanisms that are lightweight, workflow-integrated, and compatible with existing software engineering practices.

\subsubsection{\textbf{Monitoring Usability, Alert Fatigue, and Accountability}}\mbox{}\

Post-deployment monitoring introduces additional challenges beyond detecting statistical abnormalities. Leest et al. \cite{monitoring_practices_2026} show that poorly calibrated monitoring systems can generate large numbers of drift alarms and other alerts while providing limited guidance about their operational significance. Repeated non-actionable alerts can lead to \textbf{alert fatigue}, reducing practitioner trust and, in some cases, encouraging teams to ignore or disable monitoring mechanisms.

This problem extends beyond conventional drift detection. Production assurance systems may simultaneously generate data-quality warnings, outlier alerts, fairness violations, policy violations, performance degradation warnings, and security or privacy signals. However, comparatively less attention is given to the \textbf{usability and actionability} of these alerts. A detected anomaly is of limited operational value if engineers cannot determine what changed, why the change occurred, how severe it is, which pipeline component caused it, or what action should follow.

This reveals a broader observability challenge. Data engineering platforms, MLOps tools, model-serving infrastructure, and governance systems frequently monitor different parts of the pipeline in isolation. Consequently, when an anomaly occurs, practitioners may struggle to determine whether it originates from an upstream ETL failure, stale or incorrect features, model degradation, genuine population shift, or a governance-policy violation. Effective Responsible AI observability therefore requires mechanisms that connect signals across pipeline stages and provide contextual explanations rather than isolated statistical alarms.

The reviewed literature also reveals an \textbf{accountability gap}. Responsibility for AI failures may be distributed across model developers, data engineers, domain experts, deployment teams, external model providers, governance systems, and end users. When automated monitoring systems, LLM-based evaluators, or autonomous agents participate in governance decisions, responsibility becomes even more difficult to establish. Continuous assurance therefore requires stronger traceability between requirements, data, code, model versions, runtime decisions, governance evidence, alerts, and human interventions.

\subsubsection{\textbf{Documentation, Traceability, and Maturity Barriers}}\mbox{}\

Structured documentation, such as Model Cards and Data Cards, is widely recommended to improve transparency and accountability. However, the reviewed studies show that maintaining such documentation remains difficult because of organizational, workflow, and usability barriers. Pushkarna et al. \cite{data_cards_2022} show that documentation quality can suffer when responsibilities are distributed across stakeholders with different levels of technical knowledge. When documentation is created manually after development, it can quickly become outdated as the underlying data, code, model configuration, and deployment environment evolve.

This creates a \textbf{code--document drift} problem in which governance artifacts no longer accurately represent the deployed system. The problem is particularly important for Responsible AI because limitations, fairness risks, privacy assumptions, mitigation strategies, and intended-use constraints may change as the pipeline evolves. Documentation therefore needs to remain traceable to the corresponding code, datasets, experiments, model versions, and deployment configurations.

Bhat et al. \cite{DocML_2023} address part of this problem by integrating documentation directly into JupyterLab. Their DocML tool links code and documentation artifacts and prompts developers to record information while working. This illustrates the value of embedding governance activities into existing development environments rather than treating them as separate administrative tasks. Such workflow-integrated mechanisms may reduce developer effort while limiting the divergence between implementation and documentation.

Other studies emphasize progressive governance and project maturity. Cabrera et al. \cite{deblinder_2021} use end-user feedback to help developers identify previously unrecognized failure patterns, while Helmer et al. \cite{audit_catolog_case_study_2024} propose a staged MLOps maturity model with quality gates that determine whether projects are ready to progress toward production. These findings suggest that Responsible AI assurance may be more sustainable when controls are introduced progressively according to project maturity rather than imposed as a large set of independent governance requirements.

However, there remains limited guidance on which Responsible AI controls should be required at different maturity levels and how evidence from logging, documentation, testing, monitoring, and human review should be combined into a continuously maintained assurance record.

\subsubsection{\textbf{The Policy--to--Code Gap in AI Governance}}\mbox{}\

A further challenge concerns the translation of high-level governance principles into operational software controls. Responsible AI requirements are commonly expressed through abstract principles such as fairness, transparency, accountability, privacy, safety, robustness, and human oversight. However, these principles do not directly specify what developers should implement, what metrics should be monitored, or what conditions should trigger intervention. We characterize this challenge as the \textbf{Policy--to--Code Gap}.

The reviewed governance studies apply a range of mechanisms to operationalize Responsible AI requirements, including manually defined rules, policy engines, human approval loops, ontologies, knowledge graphs, and hypergraph-based representations. However, these approaches vary substantially in how policies are represented and enforced. Many still depend on manually interpreting high-level principles and encoding them into rules, thresholds, or workflow conditions.

This creates several difficulties. Governance rules can become outdated when regulations, organizational policies, models, datasets, or deployment conditions change. Different stakeholders may also interpret the same ethical principle differently, resulting in inconsistent implementations. Moreover, different governance frameworks emphasize different combinations of fairness, explainability, privacy, accountability, robustness, and human oversight, with limited agreement on how these principles should be prioritized when they conflict.

Consequently, an important open challenge is the development of systematic mechanisms for translating Responsible AI policies into \textbf{executable, traceable, and maintainable software controls}. Such mechanisms could connect high-level requirements to runtime monitors, acceptance criteria, CI/CD quality gates, documentation obligations, and escalation procedures while preserving traceability to their original governance rationale.

\subsubsection{\textbf{Delayed Feedback and Closed-Loop Assurance}}\mbox{}\

Continuous monitoring is also constrained by the availability of reliable outcome information. In many production environments, ground-truth labels are delayed, costly, or unavailable. In clinical settings, this problem is particularly important because an AI system's own recommendation may alter the subsequent outcome. For example, an intervention triggered by a high-risk prediction can prevent the adverse event that would otherwise have provided evidence about whether the prediction was correct. As a result, the system may never observe the counterfactual outcome required for direct validation.

Creating untreated holdout groups may provide counterfactual evidence, but this approach can be restricted by ethical, regulatory, clinical, or operational considerations. Consequently, continuous assurance increasingly relies on alternative feedback mechanisms.

Coombs et al. \cite{assessAI_2026} address this issue through LLM-assisted surrogate labeling for post-deployment monitoring of clinical imaging AI. Their approach extracts surrogate diagnostic labels from de-identified radiology reports and compares them with model predictions to support performance monitoring across healthcare facilities. When discrepancies occur, a local forensic workflow allows clinical administrators to inspect cases and feed validated outcomes back into the monitoring process. This establishes a pattern of \textit{closed-loop assurance}, in which automated monitoring identifies potentially problematic cases, human experts review uncertain cases, and the resulting evidence is returned to the governance system.

However, surrogate labels introduce another source of uncertainty because the mechanism generating the monitoring evidence may itself be imperfect. Continuous assurance must therefore consider not only whether the target model is reliable, but also whether the mechanisms used to generate labels, alerts, or audit evidence are sufficiently trustworthy.

\subsubsection{\textbf{Human-in-the-Loop Assurance and Organizational Friction}}\mbox{}\

Many of the reviewed studies retain humans within the governance loop, particularly when automated mechanisms cannot provide sufficient confidence. Aguilar et al. \cite{data_centric_AI_governance_2026} and Tran et al. \cite{xai_mlops_fault_detection_2024}, for example, incorporate iterative human feedback into governance and monitoring workflows, allowing domain experts to review system behaviour and intervene when automated checks are insufficient.

Human oversight is particularly important for high-risk decisions, ambiguous policy interpretations, unexpected model behaviour, and cases for which ground truth is scarce. However, relying heavily on expert review creates its own scalability and usability challenges. Continuous manual validation increases cost and cognitive workload, and different reviewers may apply different interpretations of correctness or acceptable risk. Technical teams may evaluate correctness using population-level statistical performance, whereas domain or business stakeholders may focus on whether individual outcomes cross operational decision thresholds.

The literature therefore indicates a need to move beyond simply requiring ``human-in-the-loop'' governance toward \textbf{selective human oversight}. Open questions remain regarding when human intervention should be triggered, which cases should be prioritized, who should review them, how disagreements should be resolved, and how expert feedback should be incorporated into subsequent monitoring and assurance decisions.

\subsubsection{\textbf{Non-Determinism and Emerging Assurance Challenges}}\mbox{}\

The increasing use of Large Language Models (LLMs), generative AI, and agentic workflows introduces additional assurance challenges because identical or semantically equivalent inputs may not always produce consistent outputs. Conventional test suites based on fixed input--output expectations remain useful but may be insufficient for verifying such systems throughout their lifecycle.

For neural models, Ahmed et al. \cite{deepinfer_2024} propose \textbf{DeepInfer}, which derives input-level preconditions from the internal structure of a neural network. Inputs that violate these inferred conditions can be identified as potentially unreliable before prediction. This illustrates a broader assurance approach in which verification considers not only model outputs, but also whether inputs lie within conditions where the system can reasonably be expected to behave reliably.

The problem becomes more pronounced when LLMs themselves are used as evaluators or governance components. Bena et al. \cite{LLM_certification_2026} show that LLM-based evaluation can be sensitive to the ordering of prompts and examples, producing different judgments when semantically equivalent evaluation inputs are rearranged. This creates an important \textbf{oracle reliability problem}: a governance decision may change even when the system under evaluation has not changed.

LLM-based evaluators are also subject to version changes and model drift. A prompt or evaluation procedure validated against one model version may behave differently after the provider updates the underlying model. Therefore, automated judges, compliance evaluators, surrogate-label generators, and policy interpreters must themselves be versioned, monitored, and periodically revalidated. This creates a recursive assurance challenge in which the tools used to evaluate AI systems also require assurance.

Emerging research also uses autonomous agents as part of governance workflows. Gunaweera et al. \cite{Gen_Agent_debias_2025} introduce a multi-agent framework in which specialized agents coordinate bias detection, mitigation, and re-evaluation. Although such approaches can reduce manual effort, they introduce additional uncertainty because agents may disagree, errors can propagate between agents, execution order may influence the final outcome, and responsibility for incorrect governance decisions becomes difficult to establish.

These findings suggest that assurance for generative and agentic AI must move beyond deterministic PASS/FAIL testing. More appropriate mechanisms may combine repeated evaluation, input validation, statistical confidence, independent evaluators, automated guardrails, runtime traceability, and human oversight. A particularly important unresolved problem is how to produce \textbf{stable and reproducible assurance evidence when both the system under test and the mechanisms used to evaluate it are non-deterministic}.

\subsubsection{\textbf{Summary of Operational Barriers}}\mbox{}\

Overall, RQ3 shows that the principal barriers to continuous Responsible AI assurance are not caused by a lack of proposed technical mechanisms. Instead, they arise from the difficulty of sustaining those mechanisms within operational software workflows. The literature identifies interconnected challenges involving adoption and integration overhead, limited Responsible AI logging, fragmented observability, documentation drift, alert fatigue, unclear accountability, delayed ground truth, stakeholder knowledge differences, policy-to-code translation, dependence on human review, and the non-determinism of emerging generative and agentic systems.

Collectively, these findings suggest that future Responsible AI infrastructure should move toward \textbf{workflow-integrated continuous assurance}. Such infrastructure should minimize developer effort, preserve traceability between policies and implementation, generate and maintain machine-readable assurance evidence, prioritize actionable monitoring signals, selectively incorporate human oversight, and explicitly account for uncertainty in both the AI system and the mechanisms used to govern it.

\section{Threats to Validity} 
\label{threats}
\begin{itemize}
    \item {\textbf{Search and Selection Bias}}\\
The terminology around Responsible AI and MLOps is fragmented, so relevant studies may use terms not included in the search string. This threat was mitigated through broad search terms and supplementary snowballing, but cannot be eliminated completely.

\item {\textbf{Screening Reliability}}\\
The topic has fuzzy boundaries because Responsible AI, AI governance, trustworthy AI, MLOps, and ML production overlap with many adjacent literatures. To mitigate this, the final screening prioritized operational mechanisms, ML lifecycle relevance, and evaluation maturity.

\item{\textbf{Data Extraction Subjectivity}}\\
Coding Responsible AI concerns and mechanism types requires interpretation. This was mitigated by using predefined research questions, grounded categories from prior literature, and mechanism-based themes derived from the included studies.

\item{\textbf{Evidence Maturity Assessment}}\\
Evaluation maturity is not always reported consistently. A prototype, case study, or deployment may differ across studies. The review treats maturity coding as comparative and interpretive rather than as a precise quantitative ranking.
\end{itemize}

\section{Future Directions}
\label{future}
Based on the gaps identified across the reviewed literature, several directions require further investigation: 

\begin{itemize}[leftmargin=*] 
\item \textbf{Automated Responsible AI Evidence Collection:} Develop low-friction mechanisms for automatically capturing fairness, privacy, explainability, provenance, and monitoring evidence during normal ML pipeline execution. 
\item \textbf{Developer-Centered Governance Integration:} Investigate IDE-, notebook-, and CI/CD-integrated mechanisms that reduce the documentation and configuration burden associated with Responsible AI practices. \item \textbf{Standardized Machine-Readable Compliance Evidence:} Establish interoperable formats for connecting regulatory requirements, technical controls, pipeline checks, and generated evidence across organizations and regulatory regimes. 
\item \textbf{Adaptive and Risk-Aware Governance Gates:} Develop governance mechanisms that dynamically adjust assurance requirements according to model risk, deployment context, and changes in operational conditions. 
\item \textbf{Actionable Drift Diagnosis:} Move beyond statistical drift detection toward causal and context-aware approaches that explain the source of distribution shifts and recommend appropriate operational responses. 
\item \textbf{Continuous Safety Cases and Recertification:} Investigate methods for automatically updating safety arguments and certification evidence using runtime telemetry and changes in the operational environment. 
\item \textbf{Assurance for LLMs and Agentic AI:} Develop verification and evaluation methods that account for stochastic outputs, order sensitivity, tool use, multi-agent interactions, and evolving system behavior. 
\item \textbf{Human Oversight in Continuous Assurance:} Examine how automated governance mechanisms can effectively incorporate domain experts and human approval without creating excessive operational burden. 
\item \textbf{Longitudinal Studies of Responsible MLOps Adoption:} Conduct long-term empirical studies to understand the maintenance cost, reliability, usability, and organizational adoption of continuous assurance mechanisms in production. 
\item \textbf{Bridging the Aspiration-Practice Gap:} Investigate why Responsible AI mechanisms demonstrated in research remain weakly adopted in real-world repositories and identify engineering practices that support sustained adoption. \end{itemize}

\section{Conclusion}
\label{conclusion}
In this SLR, we synthesized 44 primary studies to examine how Responsible AI principles are operationalized within MLOps pipelines and ML production workflows. The review organized the evidence around runtime observability and lineage (RQ1), codified governance and pipeline-level controls (RQ2), and the operational and non-deterministic challenges affecting their adoption (RQ3). For runtime observability and lineage, the literature analysis shows that continuous observability increasingly relies on decoupled monitoring architectures, statistical drift detection, causal diagnosis, and verifiable data and model lineage. These mechanisms allow production systems to detect behavioral changes while preserving evidence about how datasets, models, and pipeline executions evolve over time. 

For governance and pipeline-level controls, the reviewed studies demonstrate a shift from governance as retrospective documentation toward governance as an executable component of the ML lifecycle. Policy-as-Code, pipeline-native promotion gates, structured safety cases, runtime guardrails, and continuous certification mechanisms connect governance requirements directly to technical checks and deployment decisions. 

For operational and non-deterministic challenges affecting RAI adoption, however, the evidence reveals a persistent gap between proposed governance mechanisms and their practical adoption. Responsible AI telemetry, post-deployment monitoring, structured documentation, and governance-oriented tooling remain less integrated into everyday development workflows than conventional training and experiment-tracking practices. The increasing adoption of LLMs and agentic systems further complicates assurance by introducing stochastic and context-dependent behavior that cannot be fully addressed through conventional deterministic testing alone. 

Overall, the findings suggest that trustworthy ML systems require more than isolated fairness metrics, documentation templates, or pre-deployment evaluations. Responsible AI requires engineering mechanisms that can be monitored and traced throughout the ML lifecycle. These mechanisms should support governance decisions and provide clear evidence of system behavior. A key challenge for Responsible MLOps is turning these research advances into practices that can be adopted and maintained in real-world systems.


\bibliographystyle{IEEEtran}
\bibliography{references}

\appendices

\end{document}